\documentclass[reprint,
superscriptaddress,
nofootinbib,
 amsmath,amssymb,
 aps,
floatfix,
]{revtex4-2}

\usepackage{graphicx}
\usepackage{dcolumn}
\usepackage{amsmath}
\usepackage{bm}
\usepackage{xcolor}

\usepackage{verbatim}
\usepackage[english]{babel}

\PassOptionsToPackage{hyphens}{url}
\usepackage[hidelinks]{hyperref} 
\usepackage{url}

\renewcommand\vec{\mathbf}
\DeclareMathOperator{\Tr}{Tr}

\begin{document}


\title{Effective reflection mode measurement \\for hanger-coupled microwave resonators}

\author{John R. Pitten}
\email{johnrpitten@gmail.com}
\affiliation{Department of Physics, University of Colorado, Boulder, Colorado 80309, USA}
\affiliation{Department of Electrical, Computer, and Energy Engineering, University of Colorado, Boulder, Colorado 80309, USA}
\affiliation{National Institute of Standards and Technology, Boulder, Colorado 80305, USA}
\author{Nicholas Materise}
\altaffiliation[Present address: ]{QuantWare, Elektronicaweg 10, 2628 XG Delft, Netherlands}
\affiliation{Colorado School of Mines, Golden, Colorado 80401, USA}
\author{Wei-Ren Syong}
\affiliation{Department of Physics, University of Colorado, Boulder, Colorado 80309, USA}
\affiliation{Department of Electrical, Computer, and Energy Engineering, University of Colorado, Boulder, Colorado 80309, USA}
\affiliation{National Institute of Standards and Technology, Boulder, Colorado 80305, USA}
\author{Jorge Ramirez}
\affiliation{Department of Physics, University of Colorado, Boulder, Colorado 80309, USA}
\affiliation{Department of Electrical, Computer, and Energy Engineering, University of Colorado, Boulder, Colorado 80309, USA}
\affiliation{National Institute of Standards and Technology, Boulder, Colorado 80305, USA}

\author{Douglas Bennett}
\affiliation{National Institute of Standards and Technology, Boulder, Colorado 80305, USA}
\author{Corey Rae H. McRae}
\affiliation{Department of Electrical, Computer, and Energy Engineering, University of Colorado, Boulder, Colorado 80309, USA}
\affiliation{Department of Physics, University of Colorado, Boulder, Colorado 80309, USA}
\affiliation{National Institute of Standards and Technology, Boulder, Colorado 80305, USA}

\date{\today}

\begin{abstract}

Superconducting microwave resonators are used in many low-power applications, such as the study of two-level system loss in superconducting quantum devices. Fano asymmetry, characterized by a nonzero asymmetry angle $\phi$ in the diameter correction method, results from the coupling schemes used to measure these devices, including the commonly used hanger method. $\phi$ is an additional fitting parameter which contains no physically interesting information and can obscure device parameters of interest. The T-junction symmetry nominally present in these resonator devices provides an avenue for the elimination of Fano asymmetry using calibrated measurement. We show that the eigenvalue associated with the common mode excitation of the resonator is an effective reflection mode (ERM) which has no Fano asymmetry. Our analysis reveals the cause of Fano asymmetry as interference between common and differential modes. Practically, we obtain the ERM from a linear combination of calibrated reflection and transmission measurements. We utilize a three-dimensional aluminum cavity to experimentally demonstrate the validity and flexibility of this model. To extend the usefulness of this symmetry analysis, we apply perturbation theory to recover the ERM in a multiplexed coplanar waveguide resonator device and experimentally demonstrate quantitative agreement in the extracted $Q_i^{-1}$ between hanger mode and ERM measurements. We observe a fivefold reduction in uncertainty from the ERM compared to the standard hanger mode at the lowest measured power, $-160$ dBm delivered to the device. This method could facilitate an increase in throughput of low-power superconducting resonator measurements by up to a factor of 25, as well as allow the extraction of critical parameters from otherwise unfittable device data.

\end{abstract}

\maketitle

\section{Introduction} 

Superconducting microwave resonators find extensive use in low power applications, such as in conjunction with superconducting qubits as ancillary readout devices and as proxy devices to directly measure the dielectric loss tangent which often limits qubit relaxation times~\cite{PappasTLS, McRaeTLS}. The low power operating requirements of superconducting resonators limit the available coupling networks that can be used to interact with these devices; the incoming wave must be attenuated and the outgoing wave amplified. This restriction results in two main approaches: reflection mode coupling~\cite{HaozhiCal,Castellanos-Beltran2007} realized with a circulator to redirect the reflected wave, and hanger mode coupling~\cite{Ganjam, WhitePaper,Gao2008} where many resonators can be multiplexed on a single feedline. Transmission mode coupling~\cite{Petersan1998} is also used in some cases.

Rather than exhibiting strictly Lorentzian behavior as expected for an \textit{RLC} resonator, the resulting response from reflection and hanger mode coupling is an asymmetric line shape. This obscures the extraction of resonator parameters such as resonance frequency and quality factor. The diameter correction method (DCM) is the most commonly used fitting method that accounts for asymmetric line shape \cite{Khalil, Probst}. The underlying cause has been identified as interference from a parallel signal which does not scatter off the resonator, known as Fano interference \cite{Fano, Rieger}. 

Fano interference has been previously analyzed for a system with nonideal reflection mode coupling realized with a circulator~\cite{Rieger}, but a particular interfering path for a given asymmetry $\phi$ was not identified, and instead a range in uncertainty of the internal Q-factor $Q_i$ was reported in that work. Unless otherwise stated, reflection mode refers to an ideal reflection mode as in Fig.~\ref{ReflMode}, in contrast to an imperfect reflection mode implemented with a circulator. Our work focuses on hanger-mode coupling, and by exploiting the very interference which leads to Fano asymmetry in these devices, we reconstruct a symmetric line shape, in what we call an effective reflection mode (ERM) measurement.

To achieve this, we build upon previous efforts to describe the general microwave network properties of hanger-coupled superconducting resonators~\cite{GaoThesis, Khalil, Rieger}, with an emphasis on accurate separation of internal and coupling losses, as well as the origin of Fano interference in these devices. We develop a model which provides a full scattering matrix description of hanger-coupled microwave resonators, while ensuring passivity in the scattering parameters. The need for diameter correction via the DCM is eliminated in this model, yet the two methods show agreement in all resonator parameters.

Our analysis begins by employing a scattering description of microwave devices to allow the straightforward assignment of loss to coupling and internal mechanisms. Next, we distinguish the coupling network from the resonator itself. Both devices are analyzed separately before being brought together. In particular, we assume the \textit{RLC} resonator has a reflection coefficient $\Gamma$, and the coupling network has a scattering matrix $S$. Then we calculate the reduced scattering matrix $S^R$ which describes the resonant system as a whole. Two coupling networks are considered. A two-port reactive coupling network is used to define an ideal reflection mode, and a three-port T-junction is used to model a hanger-coupled device. A discrete symmetry is initially assumed in both coupling networks for a tractable analysis. Finally we use a simple perturbation theory to extend the applicability of our analysis to devices whose coupling networks exhibit an approximate T-junction symmetry.

We experimentally verify the basic predictions of our analysis with measurements of a three-dimensional (3D) aluminum cavity using an electronic calibration unit (ECal), and demonstrate how an ERM measurement allows us to identify nonideal properties in the cavity-coupling system. Following our treatment of coupling networks with perturbed symmetry, we perform cryogenic calibrated measurements on a multiplexed coplanar waveguide (CPW) resonator device. These measurements demonstrate quantitative agreement in the measured $Q_i$ between hanger mode and ERM measurements, verifying the results of our perturbation theory calculations, and suggest an ERM advantage in the measured uncertainty of $Q_i$.

This description of the coupled resonator system results in a significant leap in our understanding of the microwave scattering properties of superconducting resonators. In particular, the cause of Fano asymmetry is revealed to be interference between common and differential eigenmodes which exist in these devices. The ERM model proves itself flexible enough to enable easy identification and modeling of nonideal resonator properties which would otherwise condemn experimental data to poor fits. Additionally, the reduced uncertainty from the ERM measurement technique suggests a path toward more rapid resonator measurements at ultralow powers, enabling improved studies of two-level system loss in superconducting quantum devices.

\section{Two-Port Symmetric Coupling Network}

\begin{figure}
    \centering
    \includegraphics[width=\linewidth]{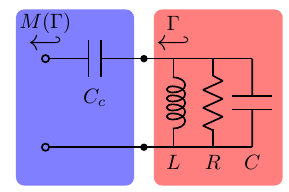}
    \caption{An idealized reflection mode coupled resonator. In red, a circuit diagram of the resonator itself, which alone has a reflection coefficient of $\Gamma$. In blue, a series reactive coupling network which modifies the resonator's reflection coefficient to $M(\Gamma)$. This mapping is used to define a reflection mode.}
    \label{ReflMode}
\end{figure}

Our construction of a reflection mode response starts by considering the microwave network behavior of a simple \textit{RLC} resonator and, separately, that of a symmetric, lossless, two-port coupling network. The combination of these devices is shown in Fig. \ref{ReflMode}. We terminate the second port of the coupling network with the resonator, and arrive at a transformation $M(\Gamma)$ from the resonator's reflection coefficient $\Gamma$ to the reflection mode response of the coupled resonator system. The reflection coefficient of a parallel \textit{RLC} resonator is 
\begin{align}\label{RLCGamma}
    \Gamma = -1 + \frac{2Q/Q_c}{1-2i\tfrac{Q}{\omega_0}\left(\omega-\omega_0\right)},
\end{align}
where $Q$ and $Q_c$ are the total and coupling Q-factors, respectively, and $\omega_0$ is the resonant frequency. 

A symmetric two-port network composed of a single series element such as a coupling capacitor can be described by a scattering matrix of the form 
\begin{align}
    \label{TwoPort}
    S = \begin{pmatrix}
        \beta & \alpha \\
        \alpha & \beta 
        \end{pmatrix}
        \quad
    \text{and}
    \quad
    \alpha + \beta = 1.
\end{align}
We rely on the fact that an $N$ port microwave network with one terminated port is an $N-1$ port network which obeys the scattering matrix reduction formula
\begin{align}\label{SR}
    S_{ij}^R = S_{ij} +\frac{S_{ik}\Gamma S_{kj}}{1-S_{kk}\Gamma }.
\end{align}
Here, $S_{ij}$ is the $N$ port network, and $S_{ij}^R$ is the reduced $N-1$ port network resulting from terminating port $k$ of the $N$ port network with a device described by a reflection coefficient $\Gamma$ \cite{Altman, PrinciplesofMicrowaveCircuits, Galwas}. The first term in Eq.~\ref{SR} represents the signal path contained in the $N$ port network, the terms in the numerator capture the signal which scatters to the terminated port $k$ as an intermediary between ports $i$ and $j$, and the denominator captures a geometric series of reflections between the terminating device and port $k$. We can apply this general formula to derive the response of the coupling network in Eq.~\ref{TwoPort} when it is terminated by a resonator. Terminating port 2 of the coupling network with the resonator results in a measured reflection coefficient of 
\begin{align}\label{RMdef}
    M(\Gamma) = \frac{\beta +(1-2\beta)\Gamma}{1-\beta \Gamma},
\end{align}
where we have used the simplifying relation $\alpha = 1-\beta$. We will take the mapping in Eq. \ref{RMdef} as the definition of a reflection mode. 
In particular, it has a frequency dependence given by 
\begin{align}\label{ReflModeFunction}
    \Gamma_{\text{RM}}(\omega) = M\left(\Gamma\left(\omega \right)\right)  = 1 - \frac{2Q/Q_c}{1-2i\tfrac{Q}{\omega_0}\left(\omega-\omega_0\right)},
\end{align}
where we have introduced the notation $\Gamma_{\text{RM}}(\omega)$ or $\Gamma_{\text{ERM}}(\omega)$ to respectively designate an ideal or effective reflection mode while emphasizing its frequency dependence over the mapping $M(\Gamma)$. Note that the values of $Q_c$ and $\omega_0$ in Eq.~\ref{ReflModeFunction} have shifted in comparison to those in Eq. \ref{RLCGamma} due to the reactive coupling (see Appendix~\ref{App:ERM-derivation}). Additionally, the off-resonant point of $1$ is a convention \cite{Khalil, Probst}. This definition of a reflection mode is a M\"{o}bius transformation of the internal reflection coefficient entirely parametrized by port reflection $\beta$ of the coupling port. M\"{o}bius transformations have the property of mapping circles to circles in the complex plane, with lines being included as a limiting case. This property ensures that the mapping $M(\Gamma$) preserves the resonant character of $\Gamma$. Additionally, any M\"{o}bius transformation can be decomposed into a combination of scaling, rotation, inversion, and translation \cite{gonzalezComplexAnalysis}. These properties are useful for interpreting and classifying transformations of the resonator response in the complex plane. For example, scaling and rotation do not affect extracted fit parameters whereas inversion and translation do affect fitted parameters. We will see an instance of translation in the complex plane once we consider perturbation theory applied to the coupling network.

Having defined the reflection mode response of a microwave resonator on a scattering basis, we now proceed with analyzing the symmetry of a T-junction. 

\section{Symmetry of the Shunt T-Junction}
\begin{figure}
    \centering
    \includegraphics[width=0.9\linewidth]{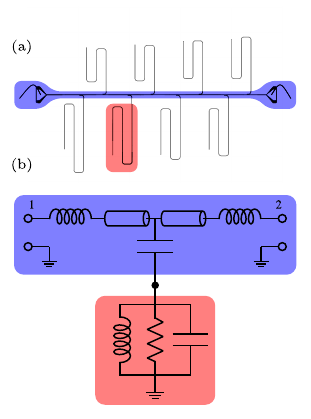}
    \caption{Schematic of a hanger coupled resonator and the associated circuit model. We find that when the coupling network adheres to a T-junction symmetry, the common mode is an effective reflection mode $S_{21}^R+S_{11}^R = M(\Gamma)$. (a) Mask for coplanar waveguide multiplexed hanger-coupled resonators \cite{WhitePaper}. Across the center is a feedline and launch pads which connect to wirebonds. These constitute the coupling network, shown in blue. In red, a quarter wave resonator. (b) Circuit diagram to model one of the resonators on this device. In blue, inductors represent wirebonds while the feedline is modeled with transmission line segments. A capacitor models the reactive coupling between feedline and resonator. Both ports share a common ground with the resonator. In red, the resonator.}
    \label{teejunc}
\end{figure}

Consider the physical construction of a hanger-coupled resonator. An example in the form of an on-chip multiplexed device is shown in Fig. \ref{teejunc}(a). There is a central feedline, off of which there are several branching quarter wave CPW resonators. At either end of the feedline microwave launch pads allow for wirebond connections which ultimately relay the signal between the chip and its package. A potential microwave circuit model for one of these resonators is shown in Fig. \ref{teejunc}(b); other models may be proposed which attempt to model more parasitics, but we will soon see that these systems are best understood not in terms of particular circuit details, but in terms of their more general properties. In our purely motivational model, the wirebonds are modeled as inductors and the feedline is a transmission line segment. Along the feedline, a reactive coupling element connects this coupling network to the resonator itself. 

Prior to labeling each circuit element of this network, we notice that the structure of the coupling network exhibits a T-junction symmetry. Of course, this is not exact: one wirebond will likely have more inductance than the other, and the two transmission line segments are certainly of different lengths. But these differences are small enough that the T-junction properties prove useful for understanding the behavior of the hanger-coupled resonator---the quantitative effects of these deviations can be accounted for later using perturbation theory. So, we are motivated to model the coupling junction not with a particular circuit, but with a more general symmetric T-junction. 

We make one crucial assumption to analyze the microwave properties of the T-junction: that it is lossless. This assumption allows us to obtain symmetry conditions via the diagonalization of the T-junction's scattering matrix,
\begin{align}
    S = \begin{pmatrix}
        \alpha & \delta & \gamma\\
        \delta & \alpha & \gamma\\
        \gamma & \gamma & \beta
    \end{pmatrix}
    \xrightarrow{\text{diag.}}\begin{pmatrix}
        s_1 & 0 & 0\\
        0 & s_2 & 0\\
        0 & 0 & s_3
    \end{pmatrix}
    ,~s_n = e^{i\theta_n}.
\end{align}

Since the T-junction's scattering matrix is unitary, its eigenvalues have unit magnitude, revealing that the T-junction has significantly fewer degrees of freedom than we initially expect. Our aim is to leverage this constraint and find symmetry conditions encoding relationships between the scattering parameters.

The broad strategy to obtain these symmetry conditions is to encode the waveguide junction symmetry in a covering operation and then use this covering operation to diagonalize the T-junction scattering matrix~\cite{PrinciplesofMicrowaveCircuits, Altman}. Doing so (see Appendix~\ref{App:Tee-Junc}), we obtain
\begin{subequations}
\begin{align}
    \alpha&=\frac{1}{4}\left(2s_1+s_2+s_3\right),\\
    \beta&=\frac{1}{2}\left(s_2+s_3\right),\\
    \gamma&=\frac{\sqrt{2}}{4}\left(-s_2+s_3\right),\\
    \delta&=\frac{1}{4}\left(-2s_1+s_2+s_3\right).
\end{align}
\end{subequations}
From these equations we can derive two symmetry conditions of the lossless T-junction~\cite{Altman}:
\begin{align}
\label{ThreePort}
    S &= \begin{pmatrix}
        \alpha & \delta & \gamma\\
        \delta & \alpha & \gamma\\
        \gamma & \gamma & \beta
    \end{pmatrix}
    \quad
    \text{and}
    \quad 
    \begin{cases}
        \alpha + \delta = \beta\\
        \beta + \sqrt{2}\gamma = 1
    \end{cases}\;,
\end{align}
where $s_3=1$ is a global phase choice analogous to $\alpha+\beta = 1$ in Eq. \ref{TwoPort}. We have now achieved our goal behind performing the symmetry analysis of the T-junction and can proceed to analyzing the behavior of a resonator which terminates the third port of a T-junction.

Examining the symmetry conditions of Eq.~\ref{ThreePort}, we see that $\alpha + \delta = \beta$ may allow us to construct an expression in which dependence on $\alpha$, $\beta$, $\gamma$, and $\delta$ is expressed solely in terms of $\beta$. Notably, the parameter $\beta$ in this system has the same physical meaning as in Eq. \ref{TwoPort}; it is the reflection coefficient of the port we intend to terminate with a resonator. Given this similarity, we may be able to construct a transformation similar to Eq.~\ref{RMdef} for a hanger-coupled resonator. Taking inspiration from this possibility, we will calculate the sum $S_{11}^R+S_{21}^R$ after terminating the third port of a T-junction with a resonator whose reflection coefficient is $\Gamma$:
\begin{align}
    S_{11}^R+S_{21}^R&=\alpha +\frac{\gamma^2 \Gamma}{1-\beta \Gamma}+\delta +\frac{\gamma^2 \Gamma}{1-\beta \Gamma}.
\end{align}
These four terms represent the ways an incident wave at port 1 will scatter. The first term represents a reflection which avoids the resonator entirely, and the second represents reflected signal which does interact with the resonator. Likewise the latter two terms represent transmitted signal. The sum $S_{11}^R + S_{21}^R$ then represents the totality of the signal scattered by the hanger-coupled resonator while maintaining phase coherence between the two ports. Upon substituting the symmetry conditions of Eq.~\ref{ThreePort}, simple manipulations lead us to
\begin{align}
    S_{11}^R+S_{21}^R = M(\Gamma) = \frac{\beta +(1-2\beta)\Gamma}{1-\beta \Gamma}.
\end{align}

Evidently, the sum $S_{11}^R+S_{21}^R$ has the same form as a reflection mode. For this reason we will call it an ERM. Immediately, we can see that a measurement of this mode confers a number of advantages. It must have zero Fano asymmetry, in contrast to a hanger-mode response or a reflection mode implemented with a circulator. It enables the abstraction of the two-port coupled-resonator system as a one-port network, which allows us to apply concepts from the simpler reflection mode circuit to the more complicated hanger-mode geometry. For example, this one-port abstraction can be used to define an effective input impedance, from which one can calculate Q-factors~\cite{PrinciplesofMicrowaveCircuits}. Additionally, we can define a plane of detuned short or open~\cite{ginzton2012microwave} which enables us to work with a vastly simplified equivalent circuit. For instance, a resonator viewed from a plane of detuned short is described by a parallel \textit{RLC} resonator as in Eq.~\ref{RLCGamma}---its off-resonant point looks like a short. On a more intuitive level, an ERM measurement will result in more information being obtained in the measurement because all available measurement power is recaptured, resulting in a signal-to-noise ratio (SNR) improvement.

Since an ERM measurement requires phase coherence between measurements at different ports, it requires two-port vector network analyzer (VNA) calibration~\cite{Ranzani}. Having demonstrated the existence of an ERM in hanger-coupled resonators, we now want a complete description of the microwave network. This would allow us to derive the hanger-mode response, and provide insight into the cause of Fano asymmetry in the hanger-mode line shape.

\section{Spectral Decomposition of the Coupled Resonator System}

Consider a symmetric T-junction terminated at port 3 by a resonator. We can construct the reduced scattering matrix using the scattering matrix reduction formula
\begin{align}
    S^R&=
    \begin{pmatrix}\label{reduced tee junction}
        \alpha + \zeta & \delta + \zeta\\
        \delta + \zeta & \alpha + \zeta
    \end{pmatrix}
    \quad \text{where} \quad 
    \zeta = \frac{\gamma^2\Gamma}{1-\beta \Gamma}.
\end{align}
It is simple to show that this reduced scattering matrix has differential and common mode eigenvectors, with corresponding eigenvalues
\begin{align}
    S_{\text{CM}} &=  \alpha + \delta + 2\zeta = \Gamma_{\text{ERM}} \quad &&\vec a_{\text{cm}} = 
    \frac{1}{\sqrt{2}}\begin{pmatrix}
        1\\1
    \end{pmatrix}\label{CM},\\[2ex]
    S_{\text{DM}} &= \alpha - \delta = -e^{-2i\phi} \quad &&\vec a_{\text{dm}} =\frac{1}{\sqrt{2}} \begin{pmatrix}
        1\\-1
    \end{pmatrix}\label{DM}.
\end{align}

Note that $S_{\text{DM}}$ is the same as the first eigenvalue of the T-junction, and therefore has unit magnitude since the T-junction is lossless. This can also be inferred from the cancellation of $\zeta$, which contains all of the dissipation in the circuit. Physically, this means that the differential input mode $\vec a_{\text{dm}}$ does not reach the resonator because of destructive interference---there is a node at the coupling junction. Conversely, the ERM has a standing wave antinode at the coupling junction resulting from constructive interference. We can use these results to derive $S_{21}^R$ in terms of the hanger-coupled resonator eigenvalues:
\begin{align}\label{fano}
    S_{21}^R = \tfrac{1}{2}\left(e^{-2i\phi}+\Gamma_{\text{ERM}}\right).
\end{align}
Here, we see explicitly the interference which causes Fano asymmetry --- it is occurring in the diagonal basis. The input signal is a linear combination of eigenvectors of the coupled-resonator system, meaning that the response is likewise a linear combination of its eigenvalues. One of these eigenvalues is the ERM response $\Gamma_{\text{ERM}}$ while the other is a phase factor. Interference between these two signals causes the asymmetry present in a hanger mode response. Having made this realization, we can proceed to obtain the hanger-mode response in a more familiar form:
\begin{align}
    S_{21}^R\left(\omega\right) &= \frac{1}{2}\left(e^{-2i\phi(\omega)}+1-\frac{2Q/Q_c}{1-2i\tfrac{Q}{\omega_0}(\omega - \omega_0)}\right)\\[2ex]
    &=e^{-i\phi}\cos \phi \left(1- \frac{\tfrac{Q}{Q_c}\left[1+i\tan \phi \right]}{1-2i\tfrac{Q}{\omega_0}(\omega - \omega_0)}\right),\label{hanger mode}
\end{align}
where $\frac{1}{Q} = \frac{1}{Q_c}+\frac{1}{Q_i}$. Here we have derived a form of the hanger mode response which can be compared to the DCM~\cite{Khalil}. Note that we have agreement (up to a complex prefactor) with the DCM under a redefinition of the coupling Q-factor 
\begin{equation}
    \frac{1}{Q_c}\rightarrow \frac{\cos\phi}{Q_c}.
    \label{eq:qc}
\end{equation}

\begin{figure*}[t]
    \centering
    \includegraphics[width=\textwidth]{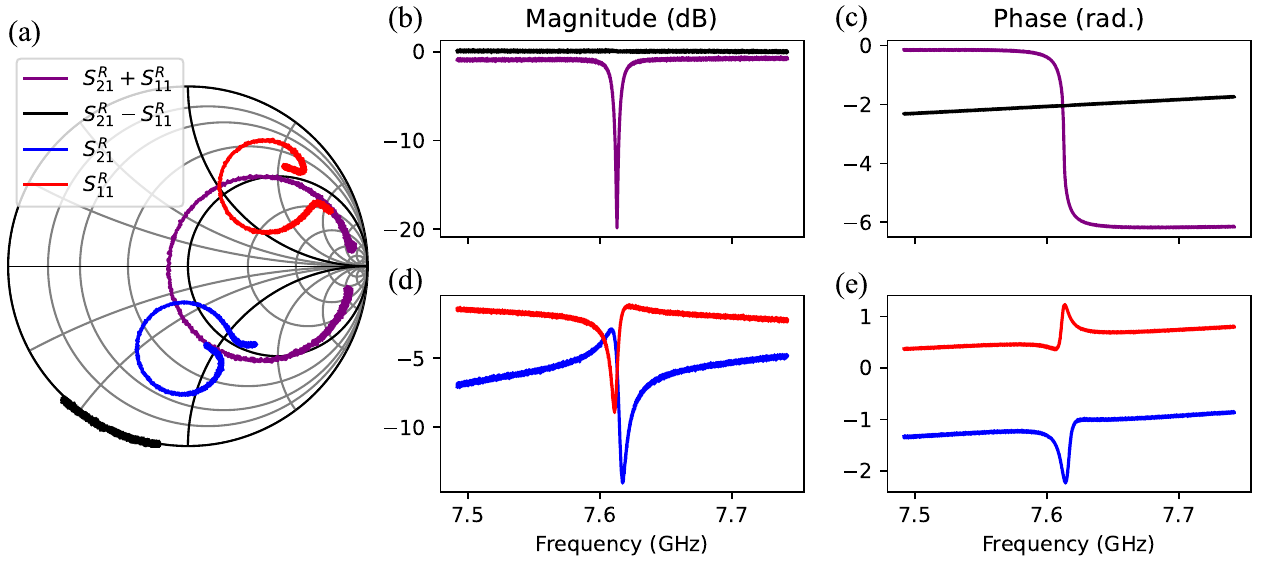}
    \caption{Experimental demonstration of an ERM measurement for a hanger-coupled 3D aluminum cavity measured at room temperature with an ECal. Smith chart (a), scattering parameter magnitude $|S|$ ((b) and (d)), and scattering parameter phase $\angle S$ ((c) and (e)) are plotted for the ERM $S^R_{21}+S^R_{11}$ (purple), differential mode $S^R_{21}-S^R_{11}$ (black), transmission $S^R_{21}$ (blue), and reflection $S^R_{11}$ (red). The ERM displays characteristics of an ideal reflection mode: a $2\pi$ phase shift and no asymmetry, unlike the reflection and transmission data. The differential mode $S^R_{21}-S^R_{11}$ displays complete destructive interference with constant unit magnitude.}
    \label{fig:CavitySummary}
\end{figure*}

Further, this model of a hanger mode makes some practical improvements over the DCM. Our definition of $Q_c$ remains the same whether the resonator is probed in a reflection mode or a hanger mode, and avoids the introduction of an imaginary conductance; as conductance is by definition the real part of admittance, the concept of imaginary conductance is a contradiction. Moreover, the fundamental definition of $Q$ is energy based and does not admit complex values~\cite{Altman, PrinciplesofMicrowaveCircuits, ginzton2012microwave}, which is consistent with our definition in Eq.~\ref{eq:qc} but not that of the DCM. Additionally, the functional form in Eq.~\ref{hanger mode} is consistent with a physical interpretation of Fano asymmetry because it is derived from Eq.~\ref{fano}. The explicit prefactor prediction of $e^{-i\phi}\cos\phi$ quantifies the SNR advantage of reflection mode over hanger mode; since $\cos\phi$ is a common factor in Eq. \ref{hanger mode} the signal is correspondingly reduced. Since the noise level remains the same, this results in a decrease in SNR. Additionally, the prefactor provides an off-resonant point which ensures $|S_{21}^R|\leq 1$, while defining a plane of detuned open \cite{ginzton2012microwave} for the corresponding ERM.

This model even allows the analysis of data otherwise too asymmetric to fit with the DCM. The $\cos \phi$ prefactor predicts the line shape turning over to form a peak rather than a dip for values of $\phi$ larger than $\pi/2$.

\section{Experimental verification with room temperature 3D cavity}

Now that we have completed the analysis of a symmetric resonator device, we can demonstrate the veracity of the ERM model. A 3D aluminum cavity measured at room temperature with a VNA and ECal is particularly well suited for this demonstration (Fig.~\ref{fig:CavitySummary}): first, because an SMA T-adapter used to couple to the device adheres to the assumed T-junction symmetry to within manufacturing tolerance and, second, because an ECal enables a highly accurate VNA calibration to eliminate the systematic errors present in the larger microwave network beyond the system of interest. This well-established performance of an ECal is in contrast to a cryogenic two-port VNA calibration, whose accuracy is much more difficult to verify~\cite{VNAHandbook}, albeit essential for a useful ERM measurement of a superconducting device.

The data in Fig.~\ref{fig:CavitySummary} confirm the main qualitative predictions of the preceding analysis. Figure \ref{fig:CavitySummary}(a) displays the complex scattering data on a Smith chart, which is simply a polar plot with impedance-based grid lines \cite{PrinciplesofMicrowaveCircuits, Altman}. The common mode $S_{21}^R+S_{11}^R$ is a reflection mode with no visible asymmetry in the line shape. A least-squares fit with a phenomenological asymmetry parameter results in an asymmetry of less than a degree. The size of the ERM circle on the Smith chart is roughly twice as big as the circles representing transmission or reflection. This difference is represented by the $2\zeta$ term in Eq.~\ref{CM} whereas the scattering parameters in Eq.~\ref{reduced tee junction} contain only a single $\zeta$. The enhanced size of the ERM resonance circle compared to that of a hanger mode suggests that the ERM has stronger coupling. In particular, a moderately undercoupled hanger mode may have an associated ERM which is critically coupled or even overcoupled. We may be able to take advantage of this change in coupling to improve resonator power sweeps, in  which the resonators are generally undercoupled at low powers to prevent overcoupling in the high-power regime. This change in coupling results from the power dependence of $Q_i$. Additionally, the differential mode $S_{21}^R-S_{11}^R$ has unit magnitude, confirming the predicted destructive interference which prevents the resonance from affecting this mode at all. The fact that its phase differs greatly from $\pi$ corresponds to the large asymmetry seen in the scattering parameters $S_{21}^R$ and $S_{11}^R$.

The ERM characterization of this Al cavity reveals additional insights that would otherwise go unnoticed. In order to preserve the device symmetry when removing the electrical delay, the same amount of delay must be removed from ports 1 and 2. Upon removing the delay from the ERM we find that the differential mode has a significant change in phase over the measurement bandwidth. This is not an arbitrary residual delay; it reflects the fact that the T-adapter scattering properties change appreciably on the scale of the resonator linewidth. Referring to Eq.~\ref{DM} and \ref{hanger mode}, we see that linear frequency dependence in the asymmetry $\phi(\omega)$ introduces two complications to the hanger mode line shape. The first is the linear background explained in the prefactor of Eq.~\ref{hanger mode}. The second is that even after implementing a background removal as suggested in Ref.~\cite{McRaeTLS}, the effects of a frequency dependent asymmetry must still be taken into account because the $1+i\tan\phi(\omega)$ term in the numerator of Eq. \ref{hanger mode} will still vary.

Examining the ERM in Fig.~\ref{fig:CavitySummary}, we see that the off-resonant point does not quite reach a magnitude of $0$ dB whereas the differential mode does. This discrepancy indicates a small amount of loss present in the T-junction's coupling arm---there, it will affect the common mode but not the differential mode. Assuming the ERM is being modeled as a series \textit{RLC} resonator, an external shunt conductance can be used to model this lossy coupling. Equation~\ref{ReflModeFunction} would simply be modified to 
\begin{align}
    s_{\text{cm}} = \left(\frac{1-g_{\text{ext}}}{1+g_{\text{ext}}}\right)\left(1 - \frac{2Q/Q_c}{1-2i\tfrac{Q}{\omega_0}\left(\omega-\omega_0\right)}\right),
\end{align}
where $g_{\text{ext}}$ is the normalized external shunt conductance.

To summarize, this example of the Al cavity confirms the qualitative predictions of the ERM model, and demonstrates that an ERM characterization of a hanger-coupled resonator allows for easy identification of nonideal properties, despite the fact that these complications cannot be clearly seen in $S_{21}^R$ or $S_{11}^R$. 

\begin{figure*}[t]
    \centering
    \includegraphics[width=\textwidth]{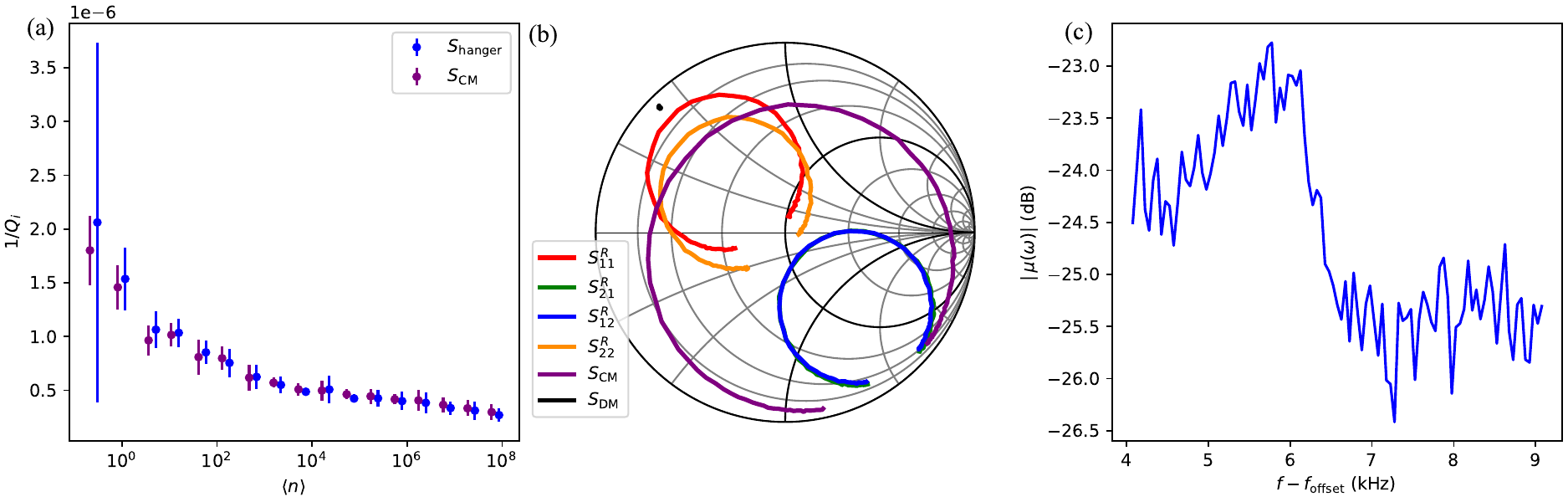}
    \caption{Experimental demonstration of an ERM measurement at $10$ mK on a multiplexed hanger-coupled coplanar waveguide resonator with perturbed T-junction symmetry. (a) Comparison of inverse internal quality factor $Q_i^{-1}$ between hanger-mode $S_{\text{hanger}}=\tfrac12(S^R_{21}+S^R_{21})$ and ERM $S_{\text{CM}} = \tfrac12(S^R_{21}+S^R_{21}) + \tfrac12(S^R_{11}+S_{22}^R)$ measurements. All values agree over a large range of powers delivered to the device, expressed in terms of average photon number $\langle n \rangle$. $S_{\text{hanger}}$ values have been slightly offset for readability. Uncertainty is reported as 95\% confidence intervals. (b) Smith chart showing the predicted splitting in reflection-type scattering parameters (red and orange). Eigenmodes $S_{\text{CM}}$ and $S_{\text{DM}} = \tfrac12(S^R_{21}+S^R_{21}) - \tfrac12(S^R_{11}+S_{22}^R)$ shown in purple and black, recovered by averaging the reflection at the two ports as prescribed by the perturbation theory calculations. Data are shown at $\langle n \rangle = 8.2\times 10^6$. (c) Magnitude of junction asymmetry $|\mu|$ obtained from scattering parameter measurements as a function of measurement frequency $f$ where $f_{\text{offset}} = 4.7076$ GHz.}
    \label{fig:CPW}
\end{figure*}

\section{Deviations from Perfect Symmetry}

In many cases it is desirable to multiplex several resonators on a single feedline. Doing so necessarily breaks the T-junction symmetry assumed in the previous analysis. However, this deviation usually remains small, so it can be effectively treated as a perturbation. We will only consider such perturbations which preserve energy conservation in the coupling junction, since the structural deviations we have in mind are not due to loss. Therefore, the perturbed coupling junction, described by $S$ can be obtained from a unitary transformation of the unperturbed junction $S_0$, which has the same form as in Eq.~\ref{ThreePort}. Expanding this unitary transformation to first order,
\begin{align}\label{tee junction perturbation}
    S = e^{-iG}S_0\, e^{iG} \simeq S_0 +\left[ S_0, iG\right],
\end{align}
where the commutator $\left[ S_0, iG \right]$ is identified as the perturbation term $\delta S$. Unitarity of both $S$ and $S_0$ requires that $G$ is a $3\times 3$ Hermitian matrix. In particular this means that the perturbation generator $G$ has a linearly independent basis in the Gell-Mann matrices $\lambda_n$. Originally introduced in the context of quantum chromodynamics \cite{Gell-Mann}, the Gell-Mann matrices generalize the Pauli matrices from $\text{SU}(2)$ to $\text{SU}(3)$. We express $G$ as
\begin{align}\label{basis for G}
    G = \sum_{n=1}^8 g_n \lambda_n,
\end{align}
where $\left|g_n\right| \ll 1$ and $\lambda_n$ is a Gell-Mann matrix. Since there are only eight, each commutator $\left[S_0, i\lambda_n\right]$ can be evaluated in turn. We find that of these eight choices, five result in nonreciprocal networks ($S_{ij}\neq S_{ji}$), indicating broken time-reversal symmetry via the introduction of a ferromagnet or topological insulator. For our purposes these terms will be discarded. Of the remaining valid generators $\lambda_2$, $\lambda_5$, and $\lambda_7$, the linear combination $\lambda_+ = \tfrac{1}{\sqrt2}\left(\lambda_5 + \lambda_7\right)$ preserves the T-junction symmetry. As such, this component is properly treated as being included in $S_0$ rather than the perturbation $\delta S$. This leaves $\lambda_2$ and $\lambda_- = \tfrac{1}{\sqrt2}\left(\lambda_5-\lambda_7\right)$ as the two independent generators of the perturbation in Eq.~\ref{tee junction perturbation}.

At this point we are ready to terminate the third port of the perturbed coupling junction with the \textit{RLC} resonator shown in Fig \ref{ReflMode}(b), and calculate the resulting perturbation of the reduced two-port coupled-resonator system. We find that a splitting is introduced in the two scattering parameters of this system which represent reflections:
\begin{align}\label{eqn:perturbed resonator system}
    S^R = \begin{pmatrix}
        \alpha + \zeta & \delta + \zeta \\
        \delta + \zeta & \alpha + \zeta
        \end{pmatrix} 
        +
        \begin{pmatrix}
            +\mu(\omega) & 0 \\
            0 & -\mu(\omega) 
        \end{pmatrix}.
\end{align}

Here, the first term has the same meaning as in Eq.~\ref{reduced tee junction}: it represents the symmetric response of the feedline-coupled resonator, and $\mu(\omega)$ contains two terms respectively proportional to $g_2$ and $(g_5-g_7)$, as defined in Eq.~\ref{basis for G}. The second term in Eq.~\ref{eqn:perturbed resonator system} represents the resulting perturbation of the reduced network. Based on this result, we see that the ERM measurement can be recovered from a general perturbation of the T-junction by simply averaging $S_{11}^R$ and $S_{22}^R$. Explicitly, 
\begin{align}
    \Gamma_{\text{ERM}} = \alpha+\delta+2\zeta = S_{21}^R +\tfrac{1}{2}\left(S_{11}^R+S_{22}^R\right).
\end{align}

Moreover, this analysis demonstrates that the transmission $S^R_{21}$ is insensitive in form to small perturbations in the coupling junction, whereas $S^R_{11}$ and $S^R_{22}$ will undergo a possibly frequency dependent displacement. These properties justify the use of Eq.~\ref{hanger mode} for multiple resonators multiplexed on the same chip, despite the fact that each resonator will see a slightly different coupling network. Additionally, there is an experimentally accessible measure of the junction asymmetry $\mu(\omega) = \tfrac12 \left(S_{11}^R-S_{22}^R \right)$, which we expect to be small but significant for relevant devices. Our approach to perturbation theory has a quantum mechanical analog in the Schrieffer-Wolff transformation~\cite{Schrieffer-Wolff}. 

\section{ERM Measurement of a Superconducting Coplanar Waveguide Resonator}

In most measurements of superconducting resonators, $Q_i$ is a parameter of strong interest. In this section, we demonstrate quantitative agreement between extracted $Q_i$ values from hanger-mode and ERM measurements on a planar CPW resonator described in Ref.~\cite{olszewski2025} (Fig.~\ref{fig:CPW}). The mask for this particular device is shown in Fig.~\ref{teejunc}(a), and design details are described in Ref.~\cite{WhitePaper}. This device was packaged and placed on the mixing chamber stage of a Janis JDry-250 dilution refrigerator. 

To implement a cryogenic thru-reflect-line (TRL) calibration, we separate the source and receivers of the VNA using its configurable test set, which is necessary to achieve VNA calibration at low powers (see Appendix~\ref{App:measurement-setup}). We use a pair of Radiall cryogenic SP6T microwave switches to cycle between calibration standards and the resonator device. The thru and line standards are realized with a pair of $3.5$mm adapters of different lengths and the reflect standards by a pair of SMA short caps. Each device is characterized at room temperature with a modern Keysight ECal and PNA. The resulting phase data is used to model the standards at cryogenic temperatures, where they are assumed lossless. An eight term calibration algorithm is implemented with the python package \verb'scikit-rf' \cite{skrf, EightTermCal}.

The cables used between the calibration plane and the actual device packaging are not phase matched, so as a preprocessing step the reference plane at port 2 of the measured CPW resonator is varied to achieve complete destructive interference in the differential mode $S_{\text{DM}}$ (see Appendix~\ref{App:preprocessing}). The resulting dataset at high power is shown in the Smith chart in Fig. \ref{fig:CPW}(b) where the differential mode appears as a black dot in the upper left. Extracted values of $Q_i^{-1}$ agree over several decades of power, shown in Fig. \ref{fig:CPW}(a), with the lowest powers suggesting an ERM advantage in uncertainty ($95\%$ confidence interval). This fivefold reduction in uncertainty seen at $\langle n \rangle = 0.3$ corresponds to a decrease in measurement time by up to a factor of 25, since SNR is inversely proportional to the square root of measurement time. 

To ensure a fair comparison between the ERM and hanger-mode measurements, the same dataset is used for both measurement modes; the hanger-mode fits simply discard the reflection data. The common mode is constructed via $S_{\text{CM}} = \tfrac12(S^R_{21}+S^R_{21}) + \tfrac12(S^R_{11}+S_{22}^R)$, where we are taking advantage of the fact that $S_{21}^R=S_{12}^R$ and averaging the transmission data to weigh transmission and reflection equally. Additionally, this averaging further suppresses noise. Similarly, the hanger-mode measurement is constructed via $S_{\text{hanger}} = \tfrac12(S^R_{21}+S^R_{21})$.

The results of our perturbation theory calculations are supported by the quantitative agreement in $Q_i^{-1}$. The Smith chart in Fig.~\ref{fig:CPW}(b) shows a significant splitting in the reflectionlike scattering parameters, and the junction asymmetry $\mu(\omega)$ is quantified in Fig.~\ref{fig:CPW}(c). Over the measurement bandwidth, $\mu$ is bound within a few decibels of $-25$ dB, which is small yet consequential. Still, we are able to recover an ERM from this dataset using the prescription developed in the preceding analysis, as evidenced by the aforementioned quantitative agreement in $Q_i^{-1}$, and the complete destructive interference seen in $S_{\text{DM}}$.

\section{Conclusion}

We have derived and experimentally verified a general microwave circuit model applicable to hanger-coupled superconducting resonators. In this model, there exists a common mode that can be treated as a reflection mode, or ERM, and a differential mode to which the resonator does not contribute due to destructive interference. These properties are agnostic to the details of the coupling network and rely only on the existence of an approximate T-junction symmetry. Losslessness is additionally assumed in the coupling network, although we have demonstrated in an example that loss can be taken into account, as well as other nonideal properties.

Additionally, we have found that the interference between the common and differential modes is the cause of the Fano interference seen in hanger-coupled resonators. In order to avoid this line shape asymmetry, the detuned common mode and differential mode should be out of phase; otherwise some nonzero asymmetry will appear. From Eqs. (\ref{CM}) and (\ref{DM}), the off-resonant point of $S_{\text{CM}} \rightarrow +1$ sets a global phase convention, and $\phi=0$ results in $S_{\text{DM}} = -1.$

On a practical level, the existence of an ERM in hanger-coupled resonators motivates the development of defined cryogenic two-port VNA calibration techniques which are independently verifiable, for example using a mismatched airline or offset shorts \cite{VNAHandbook}. These hanger-coupled resonators may even find use as calibration verification devices themselves, since the interference which characterizes their eigenmodes requires accurate measurement of both magnitude and phase.

By implementing the ERM method, resonator measurement times could be reduced by a factor of up to 25, and otherwise unfittable data sets could be accurately analyzed. The improvement of both total measurement time and ability to accurately measure nonideal devices are expected to become more important as superconducting quantum circuits scale up.

In the domain of microwave network analysis, we have demonstrated the utility of symmetry analysis applied to waveguide junctions. While the principles underlying this analysis method have been established for decades \cite{PrinciplesofMicrowaveCircuits, Altman}, modern microwave engineering textbooks make no mention of it. This omission is likely because strict adherence to the symmetry under consideration is untenable in many cases. However, we have shown through example that a Schrieffer-Wolff approach to perturbation theory may greatly extend the usefulness of symmetry analysis applied to waveguide junctions. Device symmetry may consequently be an important design consideration for the development of future superconducting quantum devices.

\begin{acknowledgments}

The contributions of J.R.P., W.-R.S., J.R., D.B., and C.R.H.M. are funded by the Materials Characterization and Quantum Performance: Correlation and Causation (MQC) program by the Laboratory for Physical Sciences. 

This work was conducted with the support of funding through the National Institute of Standards and Technology (NIST). Certain commercial materials and equipment are identified in this paper to foster understanding. Such identification does not imply recommendation or endorsement by NIST, nor does it imply that the materials or equipment identified is necessarily the best available for the purpose.

We made extensive use of the open-source python package \verb'scikit-rf' for data handling and acquisition, producing Smith chart visualizations, and for implementing VNA calibration.

We wish to acknowledge Ben Mates and Logan Howe for support through the lending of experimental equipment; Ben Mates, Florent Lecocq, and Mark Keller for helpful feedback; Maciej Olszewski for fabricating the on-chip resonator measured in this work; Suren Singh for thoughtful calibration discussions; and Rayleigh W. Parker for pointing out the similarity between our approach to perturbation theory and the Schrieffer-Wolff transformation. 

\end{acknowledgments}

\section{data availability}
    The data that support the findings of this article are openly available \cite{thedata}.

\appendix

\section{T-Junction Symmetry}
\label{App:Tee-Junc}

\begin{figure}
    \centering
    \includegraphics[width=0.45\textwidth]{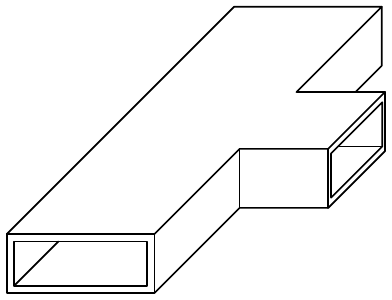}
    \caption{Illustration of a waveguide T-junction. The two collinear arms end at ports 1 and 2. These two ports are interchangeable. The coupling arm, which points to the right, ends at port 3. }
    \label{fig:tee-junc}
\end{figure}

This analysis on the symmetry of the T and other waveguide junctions is already available in the textbook references \cite{PrinciplesofMicrowaveCircuits, Altman}, but will be adapted here for completeness. Figure \ref{fig:tee-junc} shows an illustration of a waveguide T-junction. It is a lossless three-port device for which two of the ports are interchangeable. By convention these are ports 1 and 2, we will refer to port 3 as the coupling-port. The T-junction scattering matrix is

\begin{align}
    S &= \begin{pmatrix}
        \alpha & \delta & \gamma\\
        \delta & \alpha & \gamma\\
        \gamma & \gamma & \beta
    \end{pmatrix}.
\end{align}

Our goal is to find constraints among the scattering parameters $\alpha$, $\beta$, $\gamma$, and $\delta$ which result from the T-junction symmetry. We can encode this symmetry mathematically by the covering operation

\begin{align}
    F = \begin{pmatrix}
        0 & 1 & 0\\
        1 & 0 & 0\\
        0 & 0 & 1
    \end{pmatrix}.
\end{align}

Physically, this covering operation represents a reflection which exchanges ports 1 and 2 of the T-junction. Mathematically this action is represented by $F^{-1}SF = S$. Clearly $F$ and $S$ commute, so we may diagonalize $S$ by first diagonalizing $F$. This procedure will allow us to derive the symmetry conditions on $S$ from its diagonal form. 

The eigenvalues of $F$ are $f_i = \{-1, +1, +1\}$ and the corresponding eigenvectors are contained in the transformation matrix

\begin{align}
    \mathcal{F}_T = \begin{pmatrix}
        ~1 & 1 & 0\\
        -1 & 1 & 0\\
        ~0 & 0 & 1
    \end{pmatrix} = 
    \begin{pmatrix}
        \vec{x_1}& \vec{x_2}&\vec{x_3}
    \end{pmatrix}.
\end{align}

The eigenvector associated with the nondegenerate eigenvalue $f_1 = -1$ is shared by $F$ and $S$. However, the eigenvectors associated with the degenerate eigenvalues are only eigenvectors for $F$. Still, they form a convenient basis for us to find the remaining eigenvectors of $S$. A convenient choice for orthonormal real eigenvectors results in the transformation matrix
\begin{align}
\mathcal{A} = \frac{1}{2}
    \begin{pmatrix}
    \sqrt{2}&1&1\\
    -\sqrt{2}&1&1\\
    0&-\sqrt{2}&\sqrt{2}
    \end{pmatrix}
    &=
    \begin{pmatrix}
        \vec a_1&\vec a_2&\vec a_3
    \end{pmatrix},
\end{align}
where the eigenvectors satisfy $S\vec a_i = s_i \vec a_i$. Physically, this choice of eigenvectors sets the location of the reference plane at port 3 such that $s_3 = 1$. For more information consult Refs. \cite{Altman, PrinciplesofMicrowaveCircuits}. The transformation matrix $\mathcal{A}$ diagonalizes both $S$ and $F$, explicitly

\begin{align}
    \mathcal{A}^{-1}S\mathcal{A} = S_d = 
    \begin{pmatrix}
        s_1&0&0\\
        0&s_2&0\\
        0&0&s_3
    \end{pmatrix},
\end{align}

where $S_d$ is the diagonal form of $S$.
Now we can solve for $S$ and evaluate the matrix product to obtain the scattering parameters of the symmetric T-junction in terms of its eigenvalues. This procedure leads to 
\begin{subequations}
\begin{align}
    \alpha&=\frac{1}{4}\left(2s_1+s_2+s_3\right),\\
    \beta&=\frac{1}{2}\left(s_2+s_3\right),\\
    \gamma&=\frac{\sqrt{2}}{4}\left(-s_2+s_3\right),\\
    \delta&=\frac{1}{4}\left(-2s_1+s_2+s_3\right).
\end{align}
\end{subequations}
From these equations we can derive two symmetry conditions of the lossless T-junction
\begin{align}
    S &= \begin{pmatrix}
        \alpha & \delta & \gamma\\
        \delta & \alpha & \gamma\\
        \gamma & \gamma & \beta
    \end{pmatrix}
    \quad
    \text{and}
    \quad 
    \begin{cases}
        \alpha + \delta = \beta\\
        \beta + \sqrt{2}\gamma = 1
    \end{cases}\;,
\end{align}
where we have enforced $s_3=1$.

\section{Derivation of Effective Reflection Mode frequency dependence}
\label{App:ERM-derivation}

As a consequence of treating the coupling networks and resonators separately, we arrived at a natural definition of a reflection mode which takes the form of a map---more specifically, a M\"{o}bius transformation

\begin{align}\label{eqn:Mobius Transf}
    M(\Gamma) = \frac{\beta + \left(1-2\beta\right)\Gamma}{1-\beta\Gamma},
\end{align}
 from the isolated resonator reflection coefficient $\Gamma$ to the (effective) reflection mode of the coupled-resonator system $M(\Gamma)$. In order to make practical use of this construction we need to describe its frequency dependence, which we will denote $\Gamma_{\text{RM}}(\omega)$ or $\Gamma_{\text{ERM}}(\omega) = M(\Gamma(\omega))$ to emphasize the frequency dependence rather than the mapping between reflection coefficients. The subscript RM or ERM distinguishes between a conventional reflection mode applicable for resonators with a two-port coupling network and an effective reflection mode for those which couple via a T-junction. Of course, both have the same frequency dependence; the only difference is the label.

 Our derivation begins with analyzing the off-resonant point in anticipation of factoring it out of the general expression for $\Gamma_{\text{ERM}}(\omega)$. Then we put $M(\Gamma)$ in terms of $y$, the admittance of the isolated parallel \textit{RLC} resonator, and after some manipulation find the frequency dependence of $\Gamma_{\text{ERM}}(\omega)$ through that of $y(\omega)$.

 \begin{figure}
    \centering
    \includegraphics[width=\linewidth]{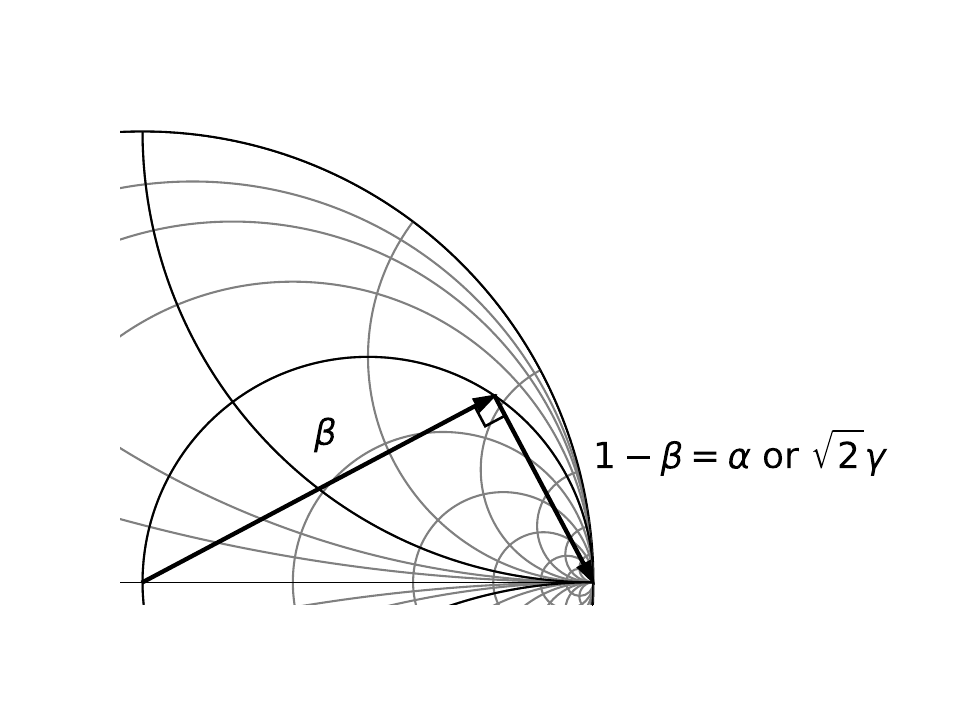}
    \caption{Geometrical construction for the phase convention of coupling-port reflection coefficient~$\beta$. In the case of a two-port coupling network $1-\beta = \alpha$ and in the T-junction $1-\beta = \sqrt{2}\gamma$. Working in terms of orthogonal variables $\beta$ and $1-\beta$ allows for simultaneous treatment of both conventional and effective reflection modes.}
    \label{beta triangle}
\end{figure}

 Since $\beta$ has the same physical meaning for both the reactive two-port coupling network and the T-junction we can treat both conventional and effective reflection modes simultaneously. In both cases we will make use of the fact that $\beta$ and $1-\beta$ are orthogonal in the complex plane. This phase condition for $\beta$ follows from unitarity of the coupling network scattering matrix, and the symmetry conditions

 \begin{align}\label{eqn:analogous symmetry conditions}
     \begin{cases}
         \beta +\alpha = 1 &\qquad\text{two-port coupling network}\\
         \beta+\sqrt{2}\gamma =1 &\qquad\text{T-junction coupling network} 
     \end{cases}.
 \end{align}

 Writing $1-\beta$ in place of either $\alpha$ for the two-port coupling network or $\sqrt{2}\gamma$ for the T-junction, unitarity imposes two simultaneous conditions

\begin{subequations}\label{eqn:unitarity conditions}
     \begin{align}
     \left| \beta\right| ^2+\left|1-\beta\right|^2 &= 1\quad\text{and}\quad\\ \beta\left(1-\beta\right)^*+\beta^*\left(1-\beta\right) &= 0,\label{inner product}
 \end{align}
\end{subequations}

 where $(\cdot)^*$ indicates the complex conjugate. The $1$ and $0$ in Eq.~(\ref{eqn:unitarity conditions}) correspond to diagonal and off-diagonal elements of $S^\dagger S$ respectively, where $(\cdot )^\dagger$ indicates the conjugate transpose. $S$ in this case can represent the scattering matrix of either the two-port coupling network or the T-junction. The second condition, Eq.~(\ref{inner product}) can be clearly interpreted as an inner product 

 \begin{align}
     \langle z, w\rangle = \frac12 \left(z\, w^* + z^* w\right) = \left|z\right| \left|w\right|\cos\theta,
 \end{align}
where $z$ and $w$ are complex numbers and $\theta$ is the difference in their arguments. Equations~(\ref{eqn:analogous symmetry conditions}) and (\ref{eqn:unitarity conditions}) then clearly show that in the case of the two-port coupling network $\beta$ and $\alpha$ are orthogonal, likewise in the T-junction $\beta$ and $\sqrt{2}\gamma$ are orthogonal. Both cases are summarized geometrically in Fig.~(\ref{beta triangle}). The reference planes of both coupling networks are chosen such that the hypotenuse of the right triangle lies on the real axis. 

Having demonstrated that $\langle\beta\,, 1-\beta\rangle=0$ holds for both the reactive two-port coupling network and the T-junction, we can proceed with deriving $\Gamma_{\text{ERM}}(\omega)$. We expect that the off-resonant point lies somewhere on the unit circle with its location determined by properties of the coupling network. Additionally we would like to rotate the off-resonant point to $\Gamma_{\text{ERM}}\rightarrow 1$, taking advantage of the fact that overall rotations do not affect fitted values of resonator parameters. This rotation can be seen as choosing a plane of detuned open as the reference plane for the ERM.

Far from resonance the reflection coefficient of the isolated parallel \textit{RLC} resonator approaches $-1$, so the off-resonant point of the reflection mode approaches 

\begin{align}
    M(-1) = \frac{-1+3\beta}{1+\beta} = \frac{2\beta -(1-\beta)}{2\beta + (1-\beta)} = \frac{\frac{2\beta}{1-\beta}-1}{\frac{2\beta}{1-\beta}+1}.
\end{align}

\begin{figure}[t]
    \centering
    \includegraphics[width=0.5\linewidth]{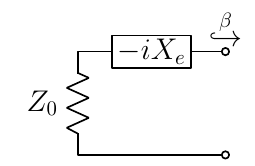}
    \caption{An equivalent circuit for the coupling network as measured at the coupling port, equally applicable to a two-port reactive coupling network or a T-junction. The absolute output impedance is $Z_{\text{out}} = Z_0 -iX_e$, resulting in a reflection coefficient of $\beta$. }
    \label{fig:coupling network equivalent circuit}
\end{figure}

Making use of the fact that $\langle 2\beta \,, 1-\beta\rangle$ = 0 we can easily see that $\frac{2\beta}{1-\beta}$ is a purely imaginary quantity. It immediately follows that the reflection mode off-resonant point $M(-1)$ lies on the unit circle, as expected. Furthermore, it will be convenient to define this imaginary quantity as

\begin{align}
    \frac{2\beta}{1-\beta} = -ix_e. 
\end{align}

In the case of reactive two-port coupling $x_e$ is the normalized reactance of the series capacitor. For the T-junction, $x_e$ can represent an effective output reactance as seen by the resonator which terminates port 3, shown in Fig.~(\ref{fig:coupling network equivalent circuit}).

With these results in mind, we will write the reflection mode mapping $M(\Gamma)$ in terms of the normalized admittance of the isolated resonator $y$. Substituting 
\begin{align}
    \Gamma = \frac{1-y}{1+y}
\end{align}
into Eq.~(\ref{eqn:Mobius Transf}) and simplifying results in

\begin{align}
    M(\Gamma(y)) = \frac{(1-\beta)+(-1+3\beta)y}{(1-\beta) + (1+\beta)y}.
\end{align}

Here we see that the off-resonant point naturally factors out

\begin{widetext}

\begin{align}
    \frac{M(\Gamma(y))}{M(-1)} = \frac{(1+\beta)(1-\beta)+(1+\beta)(-1+3\beta)y}{(-1+3\beta)(1-\beta)+(1+\beta)(-1+3\beta)y}.
\end{align}
Once again, we make the decomposition

\begin{align}
    \begin{cases}
        -1+3\beta = 2\beta -(1-\beta)\\
        1+\beta = 2\beta + (1-\beta)
    \end{cases}
\end{align}

and find 

\begin{align}
    \frac{M(\Gamma(y))}{M(-1)} = \frac{(1-\beta)\left[2\beta + (1-\beta)\right]+\left[2\beta+(1-\beta)\right]\left[2\beta-(1-\beta)\right]y}{\left[2\beta-(1-\beta)\right](1-\beta)+\left[2\beta+(1-\beta)\right]\left[2\beta- (1-\beta)\right]y}.
\end{align}

Dividing both numerator and denominator by $(1-\beta)^2$,

\begin{align}
    \frac{M(\Gamma(y))}{M(-1)} = \frac{\left(\frac{2\beta}{1-\beta}+1\right) + \left(\frac{2\beta}{1-\beta}+1\right)\left(\frac{2\beta}{1-\beta}-1\right)y}{\left(\frac{2\beta}{1-\beta}-1\right) + \left(\frac{2\beta}{1-\beta}+1\right)\left(\frac{2\beta}{1-\beta}-1\right)y}.
\end{align}

Finally, after substituting $\frac{2\beta}{1-\beta} = -ix_e$ and rearranging we find

\begin{align}\label{eqn:ginzton}
    \frac{M(\Gamma(y))}{M(-1)} = \frac{(x_e^2 +1)y+ix_e-1}{(x_e^2+1)y+ix_e+1}.
\end{align}

Now we can identify the effective input impedance as $z(\omega) = (x_e^2 +1)y(\omega)+ix_e$. Here, $y(\omega) = g(1-2iQ_i\nu)$ is the admittance of the isolated resonator, $g = Z_0/R$ is its normalized conductance, $Q_i$ is the internal Q-factor, $\nu = (\omega-\omega_0)/\omega_0$ is a normalized detuning, where $\omega_0$ is the resonant frequency. At this point we can evaluate the frequency dependence of an ERM,

\begin{align}
    \Gamma_{\text{ERM}}(\omega) = 1-\frac{2}{z(\omega) +1} = 1-\frac{2}{r(1-2iQ_i\nu)+ix_e +1},
\end{align}

where we have defined $r = (x_e^2 +1)g$, the normalized resistance of the ERM equivalent circuit as measured at the plane of detuned open. Some factoring leads to

\begin{align}
    \Gamma_{\text{ERM}}(\omega) &=  1-\frac{2}{(r+1)\left(1-2i\left(\frac{Q_i\, r}{r+1}\right)\nu +i\frac{x_e}{r+1}\right)}\\
    &=1-\frac{2/(r+1)}{1-2i\left(\frac{Q_i\, r}{r+1}\right)\nu +i\frac{x_e}{r+1}}.
\end{align}

\end{widetext}

At this point we can use 

\begin{align}
    \frac{1}{r+1} = \frac{Q}{Q_c} \text{ and } \frac{r}{r+1} = \frac{Q}{Q_i},
\end{align}
which leads to

\begin{align}
    \Gamma_{\text{ERM}}(\omega) &= 1-\frac{2Q/Q_c}{1-2iQ\nu +i\frac{x_e}{r+1}}\\
    &= 1-\frac{2Q/Q_c}{1-2iQ(\nu-\nu')},
\end{align}
where $\nu' = x_e/(2Q(r+1))$ describes the resonant frequency shift due to the reactive coupling. Practically, it is reasonable to absorb this shift in the term $(\omega-\omega_0)/\omega_0$ since we cannot experimentally distinguish $\Gamma$ from $\Gamma_{\text{ERM}}$. Finally, we have 

\begin{align}
    \Gamma_{\text{ERM}}(\omega) &= 1-\frac{2Q/Q_c}{1-2iQ\frac{\omega-\omega_0}{\omega_0}}.
\end{align}
We note that starting at Eq.~(\ref{eqn:ginzton}) our derivation closely follows that in \cite{ginzton2012microwave}.

\section{The Gell-Mann Matrices}
The Gell-Mann matrices are 

\begin{alignat*}{3}
    \lambda_1 &= \begin{pmatrix}
        0&1&0\\
        1&0&0\\
        0&0&0
    \end{pmatrix}
    \quad
    &&\lambda_2 = \begin{pmatrix}
        0&-i&0\\
        i&0&0\\
        0&0&0
    \end{pmatrix}
    \quad &~\\[1ex]
    \lambda_3 &= \begin{pmatrix}
        1&0&0\\
        0&-1&0\\
        0&0&0
    \end{pmatrix}
    \quad &~\\[1ex]
    \lambda_4 &= \begin{pmatrix}
        0&0&1\\
        0&0&0\\
        1&0&0
    \end{pmatrix}
    \quad
    &&\lambda_5 = \begin{pmatrix}
        0&0&-i\\
        0&0&0\\
        i&0&0
    \end{pmatrix}
    \quad &~\\[1ex]
    \lambda_6 &= \begin{pmatrix}
        0&0&0\\
        0&0&1\\
        0&1&0
    \end{pmatrix}
    \quad
    &&\lambda_7 = \begin{pmatrix}
        0&0&0\\
        0&0&-i\\
        0&i&0
    \end{pmatrix}
    \quad &~\\[1ex]
    \lambda_8 &=\tfrac{1}{\sqrt{3}}\begin{pmatrix}
        1&0&0\\
        0&-1&0\\
        0&0&-2
    \end{pmatrix}.\\
\end{alignat*}

They form a linearly independent basis for SU($3$) with the inner product $\frac12 \Tr \left(\lambda_m \lambda_n\right) = \delta_{m n}.$

\section{Measurement setup}
\label{App:measurement-setup}

\begin{figure*}[htb]
    \centering
    \includegraphics[width=0.85\linewidth]{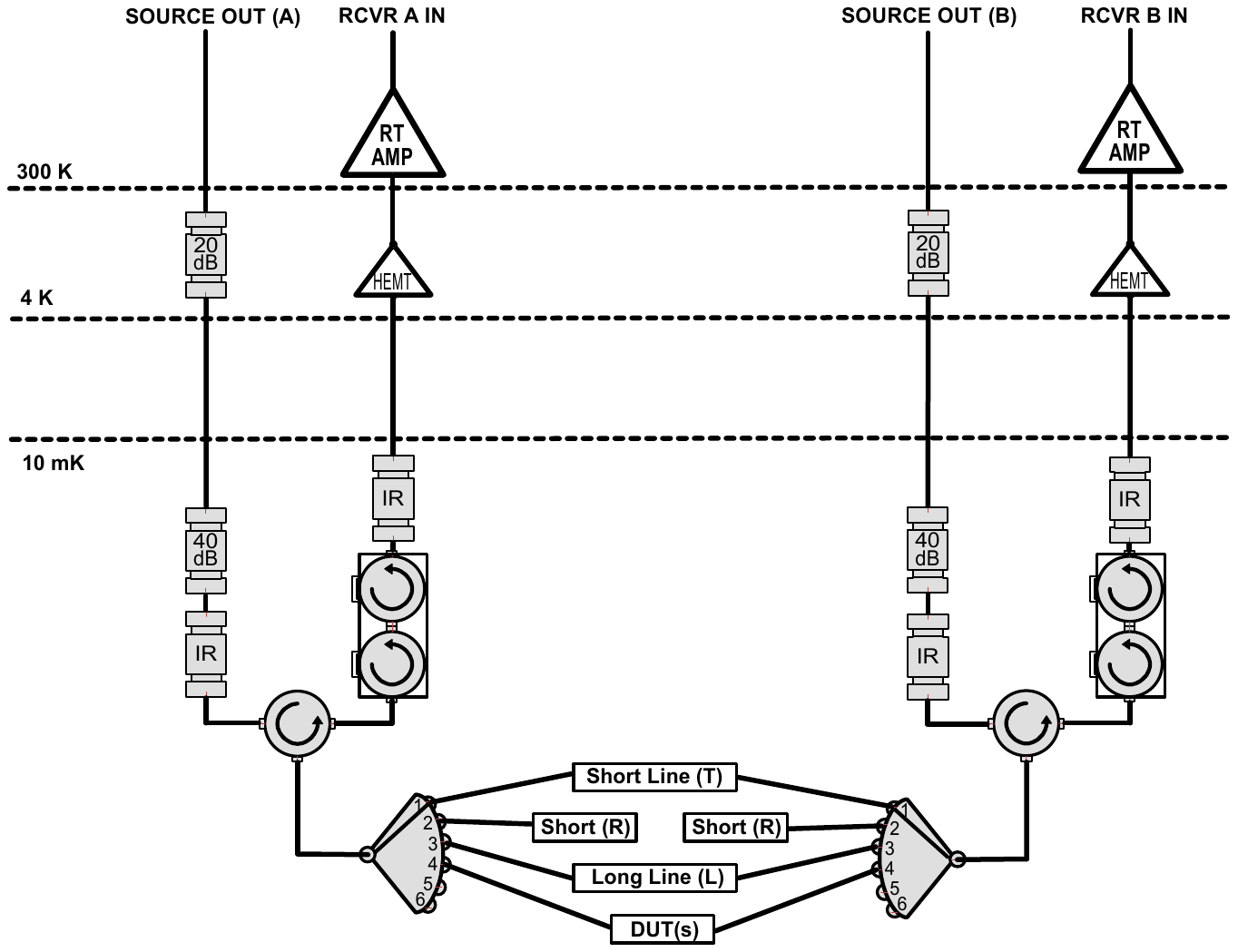}
    \caption{A wiring diagram for two-port TRL VNA calibration implemented in a dilution refrigerator. The labels ``SOURCE OUT" and ``RCVR IN" refer to the corresponding ports on the VNA's configurable test set. These ports are respectively connected to input and output lines on a dilution refrigerator. circulators at the base stage direct the microwave signals. A pair of cryogenic switches are used to cycle between calibration standards (labeled T, R, or L) and measured devices. }
    \label{fig:wiring diagram}
\end{figure*}

Figure (\ref{fig:wiring diagram}) shows the experimental setup for implementing an ERM measurement. We use the configurable test set of our VNA to separate the incident and reflected waves, using a pair of circulators to direct the incoming waves to the DUT and the scattered waves to the VNA receivers. Normally this signal separation is done using the internal VNA circuitry, but we require attenuation on the input lines and amplification on the output lines. A similar setup has been used in \cite{HaozhiCal} for one-port calibration. See \cite{ballo1997network, VNAHandbook} for more information on signal separation in VNAs. The presence of a configurable test set is a requirement for accurate low-power cryogenic VNA calibration as is necessary in this work.

At the base plate, a pair of Radiall SP6T microwave switches are used to cycle between TRL calibration standards and devices for measurement. The cables used to connect the calibration standards to the switches are all phase matched. However, the cables used to connect to the DUTs are not phase matched, necessitating an additional preprocessing step detailed in appendix~\ref{App:preprocessing}. The input lines include filtering and attenuation, the output lines include isolators, and more filtering.

Operation of the switches temporarily increases the mixing chamber stage temperature of the dilution refrigerator, however this increase is not large enough to affect TLS experiments, and a return to base temperature occurs in about 15 min \cite{HaozhiCal, Ranzani}.

\section{Protection of High-Electron-Mobility Transistor Amplifiers}
Past experience has demonstrated that the repeated, relatively frequent switch operation required of cryogenic VNA calibration is prone to damaging the high-electron-mobility
transistor amplifiers (HEMT) on the output lines. To avoid this damage we operate our HEMTs with a Keysight EDU36311A programmable power supply, ramping down the HEMT power before operating the switches, and ramping it back up prior to measurement.

\section{Preprocessing for ERM}
\label{App:preprocessing}

\begin{figure}
    \centering
    \includegraphics[width=\linewidth]{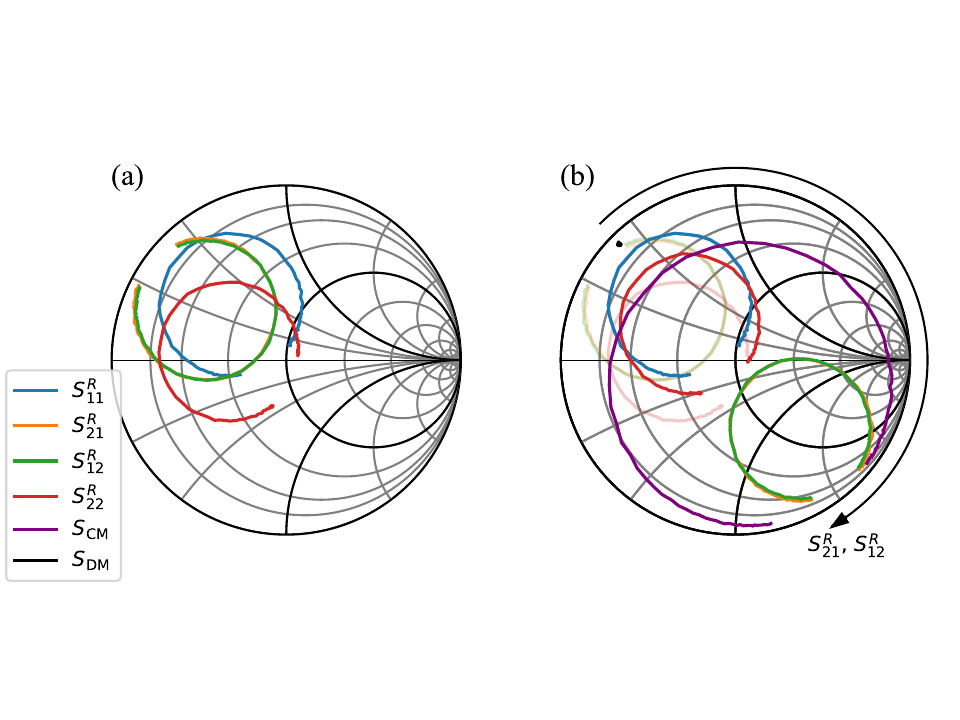}
    \caption{Illustration of the preprocessing step to account for unequal cable delay on ports 1 and 2. (a) High-power calibrated data before preprocessing. The resonant circles from transmission and reflection are misaligned in the complex plane. (b) Data after adjusting reference plane on port 2. The resonant circles are aligned to exhibit destructive interference in $S_{\text{DM}} = \tfrac12\left(S_{11}^R+S_{22}^R\right)-S_{21}^R$. The arrow around the Smith chart shows the rotation undergone by transmission scattering parameters. The reflection $S_{22}^R$ has correspondingly rotated twice the amount.}
    \label{fig:ERM preprocessing}
\end{figure}

The symmetry requirements of an ERM measurement result in an additional preprocessing step which is not required for hanger-mode measurements. The cables used to connect the resonator package to the calibration plane generally do not have the same electrical delay, so we must mathematically adjust the reference plane of one port so that delays on both ports match. A general shift in reference planes transforms the scattering matrix according to \cite{PrinciplesofMicrowaveCircuits}

\begin{align}\label{eq:cable delay}
    S' = PSP,
\end{align}

where $S$ is the original scattering matrix, $S'$ is the transformed scattering matrix, and $P$ is a diagonal matrix for which diagonal elements $P_{kk} = e^{2\pi i f \tau_k}$ represent a shift in reference plane $k$ away from the device by an amount increasing the one-way electrical delay by $\tau_k$. Note that the transformation law involves $P$ only, and not its inverse $P^{-1}$.\\

To recover an ERM measurement on a CPW resonator with perturbed symmetry we use Eq. \ref{eq:cable delay} to adjust the reference plane on port 2 such that the differential mode $S_{\text{DM}} = \tfrac12\left(S_{11}^R+S_{22}^R\right)-S_{21}^R$ exhibits complete destructive interference. Figure \ref{fig:ERM preprocessing}(a) shows calibrated data before preprocessing and Fig. \ref{fig:ERM preprocessing}(b) shows the same data after preprocessing. The arrow around Fig. \ref{fig:ERM preprocessing}(b) shows the corresponding rotation undergone by the scattering parameters $S_{21}^R$ and $S_{12}^R$, $S_{22}^R$ rotates twice as much.

\section{Average Photon Number Calculation}

Since the differential mode signal does not interact with the resonator, only the incident power contained in the common mode component should enter into the calculation of average photon number. We outline the calculation here. In the steady state, the energy contained in the resonator is

\begin{align}
    h f_0 \langle n \rangle = \frac{Q_i\, P_{\text{diss}}}{2\pi f_0},
\end{align}

where $h$ is the Planck constant, $f_0$ is the linear resonant frequency, $\langle n \rangle$ is the expected photon number, and $P_{\text{diss}}$ is the power lost to dissipation. In terms of the common mode resonator response $S_{\text{cm}}$ and the common mode power incident on the resonator $P_{\text{inc,\,cm}}$, the dissipated power is

\begin{align}
    P_{\text{diss}} = P_{\text{inc,\,cm}}\left(1-\left|S_{\text{cm}}\right|^2\right).
\end{align}

A signal $\boldsymbol{a}_{\text{inc}}$ incident on a single port of the device package contains both common and differential mode components:

\begin{align}
    \boldsymbol{a}_{\text{inc}} = \begin{pmatrix}
        1\\0
    \end{pmatrix} = \tfrac{1}{\sqrt2}\boldsymbol{a}_{\text{cm
    }}+\tfrac{1}{\sqrt2}\boldsymbol{a}_{\text{dm}}.
\end{align}

Therefore, the incident power contained in the common mode $P_{\text{inc,\,cm}}$ is half the power $P$ delivered to a port under single-ended excitation, $P_{\text{inc,\,cm}} = P/2$. The normalized dissipated power from the resonator is

\begin{align}
    1-\left|S_{\text{cm}}(f_0)\right|^2 &= 1-\left(1-\frac{2Q}{Q_c}\right)^2\\
    &= 1-\left(\frac{2Q}{Q_i}-1\right)^2\\
    &=\frac{4Q}{Q_i}\left(1-\frac{Q}{Q_i}\right).
\end{align}

Putting everything together, we have 

\begin{align}
    \langle n \rangle = \frac{P}{\pi h f_0^2}\frac{Q^2}{Q_c}.
\end{align}

In the last few expressions we have made use of 

\begin{align}
    1=\frac{Q}{Q_c}+\frac{Q}{Q_i}.
\end{align}

\section{Considerations for integration with Purcell filter qubit readout}

\begin{figure*}[!t]
    \centering
    \includegraphics[width=\textwidth]{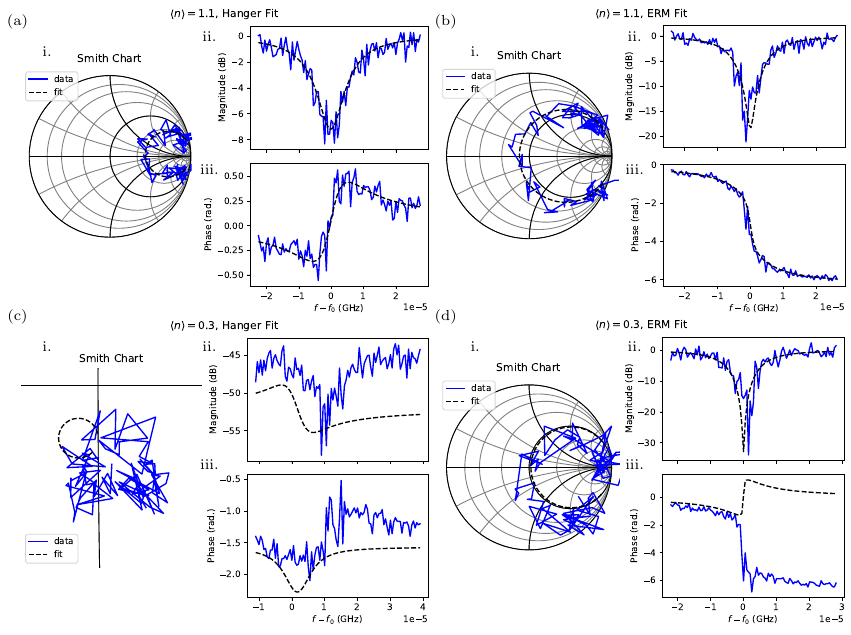}
    \caption{Fitted resonator data used in the extraction of $Q_i^{-1}$ at low powers. (a) Hanger-mode fit at $\langle n \rangle = 1.1$, (b) ERM fit at $\langle n \rangle = 1.1$, (c) Hanger fit at $\langle n \rangle = 0.3$, (d) ERM fit at $\langle n \rangle = 0.3$, (i) data and fit displayed in a Smith chart, (ii) magnitude of data and fit displayed in dB, (iii) phase of data and fit displayed in radians}
    \label{fig:fits}
\end{figure*}

One advantage an ERM measurement could confer to qubit measurement is a reconstruction of the $2\pi$ phase shift of the readout resonator [compare the $2\pi$ phase shift in Fig. 3(c) with the ``blips" in Fig. 3(e)]. If pursued, one may want to integrate this ERM qubit readout with existing devices which improve qubit performance, such as a Purcell filter, which is essentially a bandpass filter for which the readout signal is in the passband while the qubit frequency is in the stopband \cite{PurcellFilter}. Here, we briefly discuss some of the resulting considerations. 

First, one must implement VNA-like calibration with the qubit readout system. This setup would look much like the VNA calibration setup, but with the VNA sources replaced by readout signal generators, and the VNA receivers replaced by readout digitizers. The presence of the Purcell filter would certainly break the symmetry of the feedline coupled resonator. Moreover, an extended passband would come at the cost of larger insertion loss, by the Bode-Fano criterion \cite{Bode, other-Fano}. If the scattering parameters of the filter are sufficiently well-known then it can be deembedded, but this may not be feasible. If the properties of the filter can be parametrized, then it might be possible to self-calibrate by tuning these parameters until the expected destructive interference is seen on the differential mode. Another option is to introduce an identical filter on the input, but this would come at the cost of necessitating a stronger qubit drive.

\section{Comparison between Low Power fits of Hanger and ERM measurements}

Here we present the fitted data used to obtain the low power $Q_i^{-1}$ data shown in Fig. 4(a). All data was fit according to the multistep circle-fit procedure described in \cite{Probst}. Data for the second lowest power is shown in Fig.~\ref{fig:fits}(a) and (b), and data for the lowest power is shown in Fig.~\ref{fig:fits}(c) and (d). Both fits at $\langle n\rangle = 1.1$ can be considered good, but note that the ERM data displays a stronger resonance signal, with a $2\pi$ phase shift which is not present in the hanger measurement, and $20$ dB dip compared to an $8$ dB dip. 

At $\langle n \rangle = 0.3$ the advantage of the ERM over the hanger-mode measurement becomes apparent: The normalization step failed in the hanger-mode but not for the ERM. The stronger coupling in the ERM results in a doubling in the radius of the resonance circle when comparing hanger-mode with ERM. Moreover, we see critical coupling in the ERM measurement whereas the hanger measurement is so undercoupled that the noise---even after substantial averaging---obscures the data so much that the standard fitting procedure fails.

\bibliography{main.bib}

\end{document}